# Vacuum Arc Ion Sources

*I. Brown*
Lawrence Berkeley National Laboratory, Berkeley, CA 94720, USA

**Abstract**
The vacuum arc ion source has evolved into a more or less standard laboratory tool for the production of high-current beams of metal ions, and is now used in a number of different embodiments at many laboratories around the world. Applications include primarily ion implantation for material surface modification research, and good performance has been obtained for the injection of high-current beams of heavy-metal ions, in particular uranium, into particle accelerators. As the use of the source has grown, so also have the operational characteristics been improved in a variety of different ways. Here we review the principles, design, and performance of vacuum arc ion sources.

## 1    Introduction

The defining characteristic of the vacuum arc ion source is its production of *high-current* beams of *metal* ions. The metal plasma from which the ion beam is formed is produced by a vacuum arc plasma discharge, and hence the name. Vacuum arc ion sources have been used primarily for ion implantation in material surface modification research, and also for particle accelerator injection, as well as for other fundamental and applied purposes. Beams have been produced from over 50 of the solid metals of the Periodic Table (i.e., almost all of them) and carbon, with extracted ion energy up to several hundred kiloelectronvolts and beam current up to several amperes. The source is usually operated in a repetitively pulsed mode, with pulse length typically a few hundred microseconds and repetition rate up to a few tens of pulses per second; operation of a simplified d.c. embodiment has been demonstrated. In general, the ions have low but multiply ionized charge states. The mean charge state lies between 1+ and 3+ depending on the particular metal, and the charge states can be increased by up to a factor of two in a number of different ways.

The history of vacuum arc ion sources extends back to the 1940s, when attempts to use this approach for isotope separation were first made in the USA as part of the Manhattan Project during World War II [1]; the effort was abandoned, however, for reasons having to do with arc instability. Research in the general area of vacuum arcs has a long and rich history in the Former Soviet Union (FSU) [2], where the development of vacuum arc-based ion sources began in the late 1950s and early 1960s, led primarily by Plyutto and co-workers at the Sukhumy Physical Technical Institute in Georgia [3]. This early work was largely a demonstration of the basic concept of metal ion extraction from vacuum arc plasma. In the mid-1960s a kind of vacuum arc ion source was developed by workers in Ukraine for the production of beryllium ion beams with current up to 170 mA [4]. In 1979 Prewett and Holmes developed a carbon vacuum arc ion source at the University of Liverpool, UK, to produce a low-energy $C^+$ ion beam current of up to 0.5 A [5]. In the early 1980s Humphries and co-workers at the University of New Mexico, USA, embarked on an extensive investigation of the vacuum arc for production of ion beams for heavy-ion fusion research application [6–8]. A vacuum arc ion source development programme was initiated at the Lawrence Berkeley National Laboratory (LBNL) in 1982 for the production of high-current uranium ion beams for injection into the LBNL heavy-ion synchrotron (the Bevalac) for fundamental heavy-ion nuclear physics research, and later for ion implantation application. The LBNL sources, called *Mevva* (metal vapour vacuum arc) ion sources, were developed in a number of different directions, including embodiments with multiple-cathode assemblies, very large diameter extractors, miniature sources, and a test d.c. version [9, 10]. At virtually the same time, development of vacuum arc ion sources was initiated at the High Current Electronics Institute (HCEI) of the Russian Academy of Sciences, Tomsk. These sources (called *Diana* [11] and *Titan* [12]) played a significant role in the rapid growth of activity in the field throughout the FSU. A series of vacuum arc ion sources called *Raduga* were developed at the Nuclear Physics Institute (NPI) of the Tomsk Polytechnic University [13, 14], and the *Tamek* sources were developed at the Tomsk Institute of Automatic Control Systems and then at the Applied Physics Institute, Sumy, Ukraine [15]. Programmes were subsequently established in other world universities, institutes, and laboratories. A historical review of the early development of vacuum arc ion sources has been given elsewhere [16].

Vacuum arc ion source performance has been improved in a number of important ways over the years. One critical beam parameter, for example, is the ion charge state spectrum; for many purposes there is a need for higher charge states than normally formed in the vacuum arc plasma. Following this need, several different approaches to increasing the ion charge state have been developed. Beam noise and pulse-to-pulse reproducibility are also very important for particle accelerator application, and these characteristics have been improved vastly in work carried out at the Gesellschaft für Schwerionenforschung (GSI), Darmstadt, Germany. The vacuum arc ion source is now routinely used for high-current metal ion injection into the GSI heavy-ion accelerators. Reliable and long-lifetime arc triggering is another important source feature, and good progress has been made in this direction also.

Here the source performance and beam characteristics are summarized, typical design features outlined, and some of the facilities that have been established at several laboratories around the world are described.

## 2    Plasma physics of the vacuum arc discharge

Reviews of vacuum arc plasma discharges have been given by a number of authors [17–19]. The vacuum arc is a high-current discharge between two electrodes in vacuum. At the cathode the current is concentrated at a small number of tiny, discrete sites, called cathode spots. The formation of cathode spots is a fundamental characteristic of the vacuum arc discharge. The spots are where the metal plasma is produced, and it is this plasma that provides the current path between cathode and anode that keeps the arc alive. Thus some of the plasma that is generated at the cathode must necessarily deposit on the anode so as to form the cathode-to-anode current path, and some of the metal plasma can be taken away, using suitable geometry, and used for another purpose – for example, to form the ion beam of an ion source. At a cathode spot the cathode material is heated, vaporized, and ionized into the plasma state. The spot has a diameter of the order of a micrometre, and the current density is extremely high, of the order of $10^6$–$10^8$ A cm$^{-2}$. The arc current is constricted to a small number of such spots. Most of the parameters of the vacuum arc plasma are determined by the plasma physics within the cathode spot. Individual spots move around on the cathode surface, and the lifetime of a particular spot may be only microseconds or less; on the other hand, small surface irregularities like edges or protuberances tend to anchor the spots to these sites. The plasma pressure within a cathode spot is high, and the strong pressure gradient causes the plasma generated there to plume away from the surface in a manner quite similar to the plasma plume generated by an intense focused laser beam at a solid surface. The current carried by a cathode spot is typically about a few to a few tens of amperes, depending on the metal, and if the arc is caused to conduct a higher total current, then more cathode spots are formed. Thus a typical metal vapour arc discharge of several hundred amperes current might involve the participation of several tens of cathode spots. The assemblage of cathode spots gives rise to a dense plasma of cathode material that streams away from the cathode as a jet. The ion streaming velocity, or drift speed, is typically 1–3 cm μs$^{-1}$.

Ambient gas is not essential to the discharge. Much vacuum arc research has been carried out in the $10^{-6}$ Torr range, and $\sim 10^{-4}$ Torr might be considered a rough upper limit. Residual gas can have a significant influence on the plasma parameters, especially on the metal ion charge state distribution [20].

The arc current can be anywhere in the range from several tens of amperes up to many kiloamperes, and for most vacuum arc ion sources the current is typically one hundred to several hundred amperes. The arc voltage (when the arc is alight, i.e., the *burning voltage*) lies in the range 10–30 V, say typically about 20 V, and varies with the cathode metal used. Note that the arc current cannot be decreased to arbitrarily low value; the cathode spot requires a certain minimum current to stay alight, typically a few to a few tens of amperes depending on the metal species, below which the spot extinguishes. This has significance for the minimum power level at which the vacuum arc can be operated and thus on the concern of a d.c. (continuous, or steady-state) vacuum arc ion source. Essentially all vacuum arc ion sources are operated in a repetitively pulsed mode, although a d.c. version has been demonstrated.

Conduction of the arc current is supported by the metal plasma that is evolved from the solid electrode (cathode) material itself. In most vacuum arc ion sources, triggering of the vacuum arc has been accomplished by initiation of a surface flashover across a thin insulating (ceramic) surface between the cathode and an annular trigger electrode, and other methods have also been used [21].

A well-established basic feature of the vacuum arc is the relationship between the metal ion flux that is generated and the current that drives the arc. The arc current is composed of both an electron component and an ion component. The plasma ion current is the plasma flux generated at the cathode. Over a wide range of conditions the plasma ion current is a constant fraction of the arc current,

$$I_{\text{ion}} = \varepsilon I_{\text{arc}}, \tag{1}$$

where $\varepsilon$ is in the range 0.06–0.12, say typically about 10%. Thus the electrical efficiency of the vacuum arc discharge – the ratio of total generated metal ion plasma current to arc current – is high, about 10%. The electrical efficiency of the vacuum arc ion source can also be high, although the steps (total metal plasma generated by the arc) to (plasma transported to the beam formation electrodes) to (formation of ion beam from plasma presented to the extractor) to (ion beam delivered to a downstream target or beam-line) conspire together to yield an overall electrical efficiency that is considerably less than the fundamental vacuum arc discharge efficiency of 10%.

The relationship between the mass evolved from the cathode and the arc parameters has also been investigated. Generalizing, one can say that the mass of the plasma generated by the vacuum arc is of the order of several tens of micrograms per coulomb of arc current, where the precise value depends primarily on the metal used. At the cathode spots, cathode material is converted into metal plasma and also into solid *macroparticles* (so called because they are macroscopic compared to plasma particles, the ions). Macroparticles are metallic globules that are ejected from the cathode in the molten state and then rapidly solidify; they typically have a diameter in the range 0.1–10 μm. For many cases (e.g., for high-melting-point materials), the macroparticle content of the ion beam is small and is not a concern.

## 3   The basic vacuum arc ion source

The basic operating principles of the vacuum arc metal ion source can be described using the example of the LBNL Mevva II ion source [10], a simplified schematic of which is shown in Fig. 1 and its associated electrical schematic in Fig. 2. As for any plasma-based ion source, there are two basic parts – the plasma generator, and the beam extraction system. The cathode of the vacuum arc is a simple cylindrical rod (typically 5–10 mm diameter) of the material of interest, and the discharge is triggered by a high-voltage pulse applied to a trigger electrode that surrounds the cathode coaxially, separated from it by a thin (about 0.5 mm) alumina insulator. A part of the metal plasma created at the cathode flows through the anode hole, of diameter about 1 cm, and then through a drift space of several centimetres to the extractor grids. That part of the plasma that strikes the anode carries the current that keeps the plasma alight, while the plasma plume that streams through the central hole in the anode is used to form the ion beam.

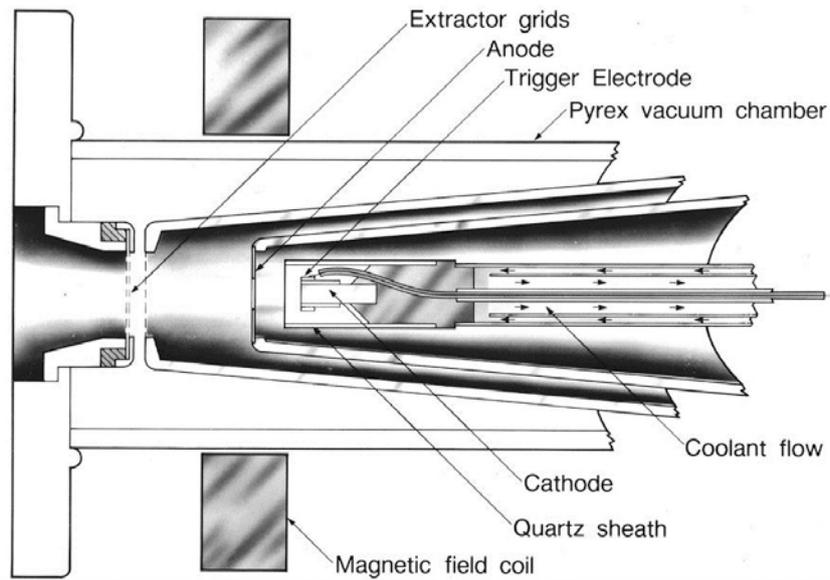

**Fig. 1:** Simplified schematic of the LBNL Mevva II ion source

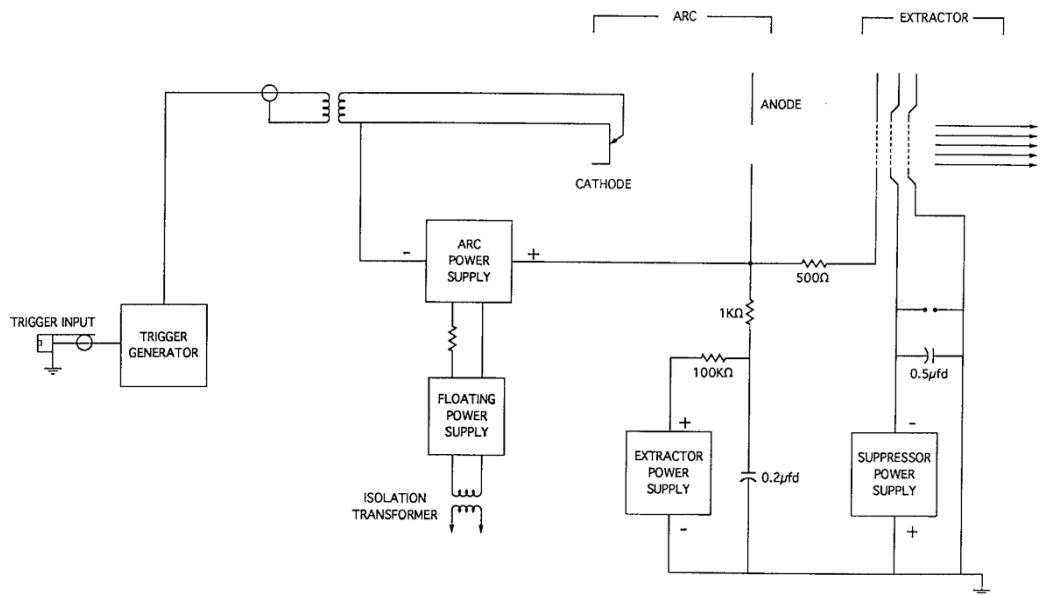

**Fig. 2:** Simplified electrical schematic of a vacuum arc ion source set-up

The kind of beam formation electrodes (*extractor grids*) used can be chosen to suit the particular application, and a conventional multi-aperture, accel–decel configuration is simple and effective. A set of three circular grids is located appropriately with respect to the expanding plasma plume, with the first grid (nearest the plasma) (*plasma grid*) at or near plasma potential, the middle grid (*suppressor grid*) typically several kilovolts negative with respect to ground so as to suppress the backflow of low-energy electrons, and the third grid (*ground grid*) at ground potential. The first grid and the entire plasma generation part of the source are biased to high positive potential. The magnetic field coil shown in Fig. 1 is optional and is often omitted. A modest longitudinal magnetic field in the arc region can be used to control the amount of plasma that is transported from the cathode to the extractor. In the absence of magnetic field, much of the plasma is lost to the walls. With an applied field of just a few hundred gauss, the radial plasma loss can be much reduced, and a greater fraction of the plasma that is formed can be presented to the extractor and converted into ion beam.

Usually the source is operated in a repetitively pulsed mode, with pulse width 100–500 μs and repetition rate up to several tens of pulses per second. The arc current is typically in the range of 50–500 A, often conveniently supplied by an *LC* (inductance-capacitance) pulse line. The trigger pulse is usually of O/C amplitude ~10 kV and duration ~10 μs or more, supplying a peak current of a few tens of amperes trigger-to-cathode by a step-up transformer, which also serves for high-voltage isolation. The ion beam extraction voltage in typical operation is several tens of kilovolts up to a maximum of about 100 kV and might be either d.c. or pulsed (d.c. in the schematic shown).

The extracted metal ion beam current can easily be as high as several amperes, and for very broad beam sources up to several tens of amperes [10], but for usual arc parameters and extractor sizes (initial beam diameter about 1–10 cm) a typical beam current might be 100–500 mA.

Beam divergence is determined primarily by the extraction optics, and if the extractor grids have been well designed and fabricated, the extraction optics can be empirically matched to the plasma by variation of plasma density via the arc current. For optimum extraction conditions (the *perveance match* condition), the beam divergence is typically about 3° half-angle for uranium.

The vacuum arc ion source produces ions that are multiply ionized, and the charge state spectrum of the ion beam is important for most applications. The ions generated in the vacuum arc plasma are in general multiply stripped, with a mean charge state of 1+ to 3+, depending on the particular metal species, and the charge state distribution can have components from $Q$ = 1+ to 6+; thus the ion energy is greater than the extraction voltage by this same factor.

## 4    Source performance

The source performance is determined in part by the specific source embodiment used, but one can nevertheless generalize, referring to specific source embodiments for specific experimental results. The operation and performance of the LBNL Mevva IV source have been described in detail [10, 22–24]. Summarized here are beam parameters that are typical of vacuum arc ion sources. Note that in all cases the source is designed and operated repetitively pulsed, not d.c. Usually the beam pulse width is a few hundred microseconds and the repetition rate, say, 10 pps. Thus this kind of ion source is suited to synchrotron injection but not cyclotron.

### 4.1    Beam current

Depending on the extractor size and thus the fraction of plasma presented for beam formation, the ion beam current can easily be as high as several amperes, and for very broad beam sources up to several tens of amperes [10]. The fraction of this beam that is usable, however – for example, for injection into an accelerator low-energy beam-line or for implantation of a downstream target – depends on the particular set-up. The beam transported downstream depends on the beam emittance and the target (or beam-line) acceptance as well as losses in the beam transport system. For example, a 10 cm extractor source version (Mevva V) has produced a titanium ion beam current into a nearby large-area Faraday cup of up to 3.5 A at 90 kV extraction voltage [10]. In another series of measurements, a source embodiment (Mevva IV) with a 2 cm diameter extractor was used to characterize the beam extraction, as shown in Figs. 3 and 4. Figure 3 shows the Ti beam current as a function of arc current for a range of extraction voltages, and Fig. 4 shows the Ta beam current as a function of extraction voltage for a range of arc currents. Measurements of this kind have been made for a number of different cathode materials (ion beam species), and good agreement was found with the well-known Child–Langmuir prediction for extracted ion current under space-charge-limited conditions,

$$I = \tfrac{4}{9}\varepsilon_o S \left(\frac{2eQ_i}{M_i}\right)^{\tfrac{1}{2}} \frac{V^{\tfrac{3}{2}}}{d^2} = 1.72\, S \left(\frac{Q_i}{A}\right)^{\tfrac{1}{2}} \frac{V^{\tfrac{3}{2}}}{d^2} \qquad (2)$$

where $S$ is the extractor open area, $Q_i$ the ion charge state (ion charge $q = eQ$), $M_i = A m_u$ the ion mass (where $A$ is the atomic weight in amu and $m_u$ the mass of 1 amu), $V$ the extractor voltage, and $d$ the extractor gap. In the second expression, $I$ is in mA, $V$ in kV, $S$ in cm$^2$, and $d$ in cm. The implication is that the ion beam current that can be expected from any embodiment of vacuum arc ion source can be predicted with reasonable accuracy from the Child–Langmuir equation above.

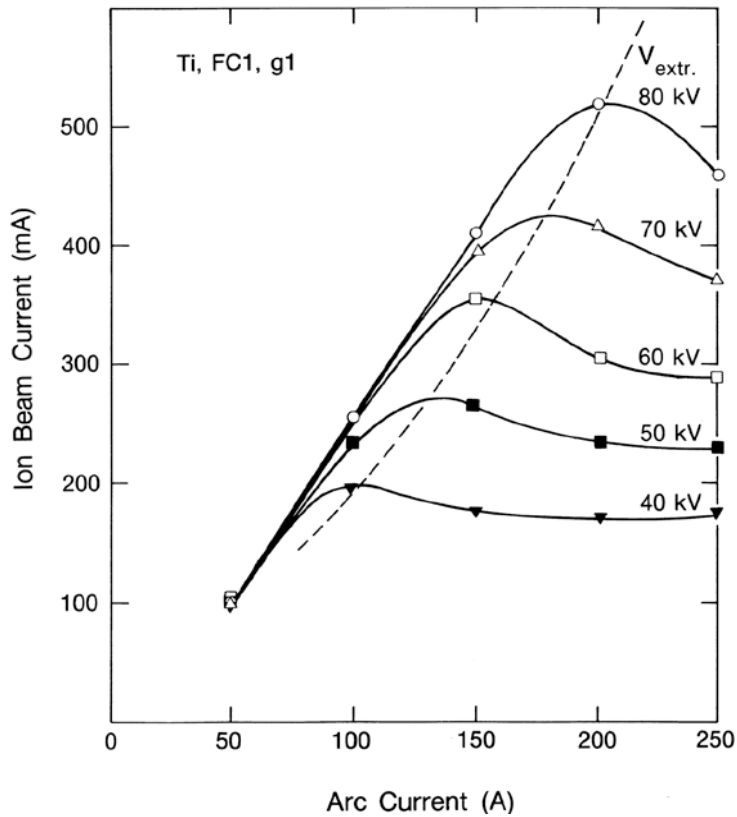

**Fig. 3:** Ion beam current as a function of arc current; titanium beam (LBNL Mevva IV)

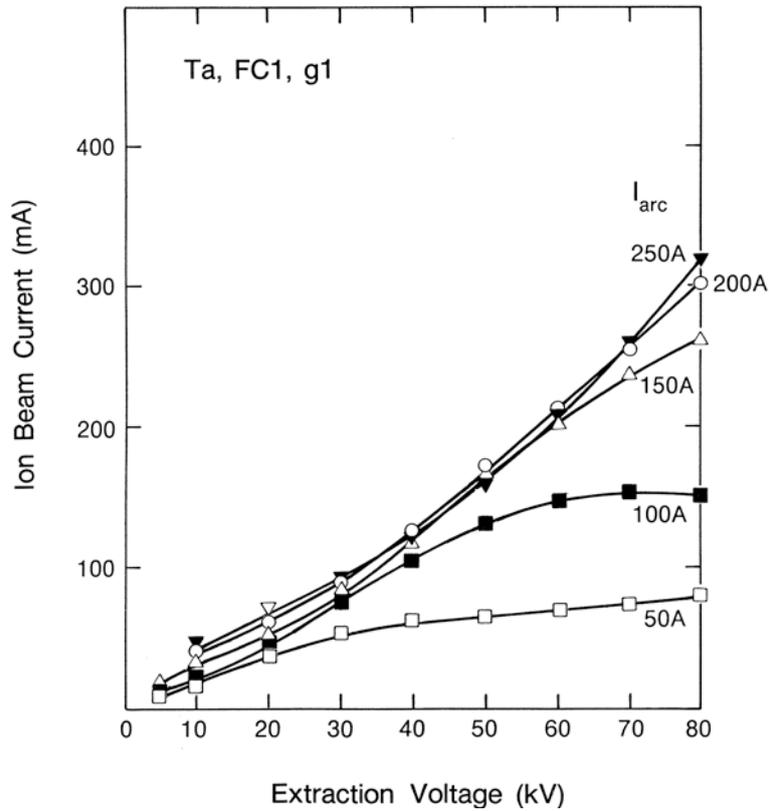

**Fig. 4:** Ion beam current as a function of extractor voltage; tantalum beam (LBNL Mevva IV)

A source embodiment has been demonstrated [10] incorporating a 50 cm diameter extractor. This source has formed Ti ion beams at 50 kV extraction voltage, i.e., about 100 keV mean ion energy (the mean charge state for Ti is 2.1+), and with peak total extracted ion current over 20 A.

### 4.2 Beam profile, divergence, and emittance

Beam divergence is determined primarily by the extraction optics, and, assuming that the extractor grids have been designed and fabricated reasonably, the extraction optics can be empirically matched by variation of the plasma density via the arc current. At optimum (the *perveance match* condition), the beam divergence is typically about 3° half-angle. The normalized emittance, $\varepsilon_N$, is about 0.3 π mm mrad (normalized) for uranium at perveance match. The ion temperature in the vacuum arc plasma is quite generally about 1 eV, and the plasma-temperature-determined beam divergence half-angle $\theta = v_\perp/v_\parallel = \sqrt{(T_{i\perp}/E_i)} = 0.2°$ for $E_i = 100$ keV. Thus the measured beam divergence, or emittance, is indeed determined mostly by the extraction and not by the transverse plasma temperature $T_{i\perp}$.

The initial beam shape, i.e., the beam radial profile immediately after extraction, is determined by the radial plasma density distribution at the extractor location together with the extraction optics. If the plasma density is not uniform across the extractor, then the extraction optics cannot be matched everywhere. An open-ended multipole magnetic bucket configured of rare-earth magnets and located in the ion source drift region can be used to flatten the plasma profile at the extractor [25], but, even so, the beam profile loses memory of its shape at the extractor fairly rapidly, and after some tens of centimetres downstream propagation the beam profile reverts to the usual Gaussian [26]. Experimentally, control of the arc current provides a convenient means of optimizing the extraction, via the plasma density. Alternatively, the extraction optics can be tailored by the extractor voltage for a fixed arc current (plasma density).

## 4.3 Beam composition

The vacuum arc ion source produces ions that are multiply ionized, and the charge state spectrum of the ion beam is important for most vacuum arc ion source applications. Ion charge state distributions (CSD) have been studied in detail experimentally [27], using time-of-flight (TOF) charge state diagnostics [28], and theoretically [29]. Almost all of the solid metals of the Periodic Table have been successfully used in vacuum arc ion sources, as well as cathodes made from metallic alloys, compounds, and pressed mixtures. Compound cathodes produce ions of the cathode constituents, and it is interesting to note that beams containing non-metallic elements, such as B and S, can be made by using conducting compound electrodes of which the non-metal is a constituent such as $LaB_6$ or $PbS$. An example of the charge state spectra obtained is shown in Fig. 5, where an oscillogram of an iridium TOF spectrum is shown (current measured by a magnetically suppressed Faraday cup). Note that for multiply charged ions, electrical current $i_{elec}$ is not the same as particle current $i_{part}$, since each particle can carry multiples, $Q$, of the electronic charge, $e$: $i_{elec} = Qi_{part}$. This can be important, as, for example, in the case of the beam current measured by a Faraday cup being electrical current, while the current needed for estimating ion implantation dose is particle current.

The mean charge state of the ion charge state distribution in general lies in the range from 1+ to 3+, depending on the particular metal species, and the distribution can have components from $Q = 1+$ to 6+; thus the ion energy is greater than the extraction voltage by this same factor. The measured charge state distributions and mean charge states for a wide range of elemental species, under typical vacuum arc operational parameters and in the absence of techniques to enhance the ion charge state, are given in Table 1. More detail is given elsewhere [27].

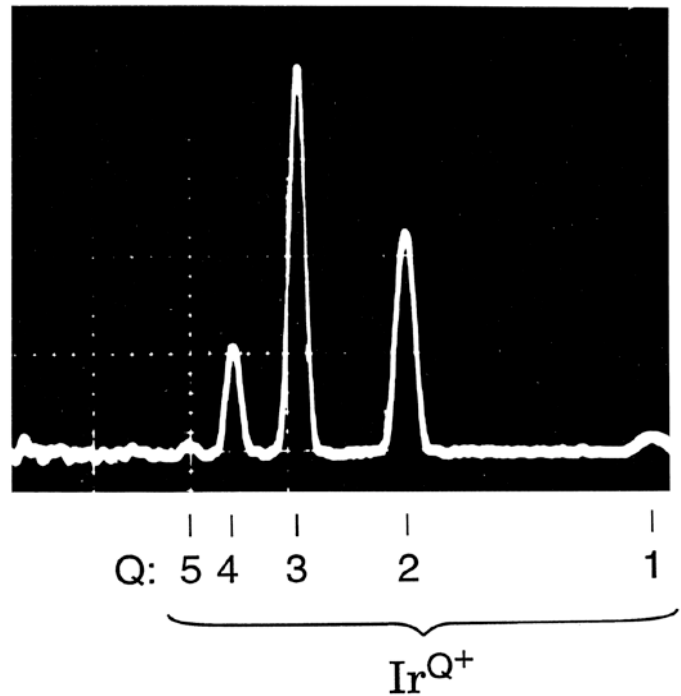

**Fig. 5**: Time-of-flight charge state distribution for an iridium ion beam. The amplitudes are electrical current measured by a Faraday cup.

## 4.4 Beam noise and pulse shape reproducibility

The basic mechanism of vacuum arc plasma formation is via explosive emission from microscopic sites on the cathode surface [30], leading to a relatively high fluctuation level in plasma density and thus also in extracted ion beam current. That is, the vacuum arc ion source beam is inherently a relatively noisy beam, and the pulse-to-pulse reproducibility is not high. While for some applications, such as broad-beam ion implantation for material surface modification, this is not a concern, for other applications, such as for particle accelerator injection, it can be a severe drawback, and for this purpose it is highly desirable to develop methods for forming a quieter beam and improved pulse shape reproducibility.

**Table 1:** Ion charge state fractions and mean charge states, expressed in terms of particle current. Asterisks (*) indicate a trace (under 1%).

| Element | Z | Charge state | | | | | | $\langle Q \rangle$ |
|---|---|---|---|---|---|---|---|---|
| | | 1+ | 2+ | 3+ | 4+ | 5+ | 6+ | |
| Li | 3 | 100 | | | | | | 1.0 |
| C | 6 | 100 | | | | | | 1.0 |
| Mg | 12 | 46 | 54 | | | | | 1.5 |
| Al | 13 | 38 | 51 | 11 | | | | 1.7 |
| Si | 14 | 63 | 35 | 2 | | | | 1.4 |
| Ca | 20 | 8 | 91 | 1 | | | | 1.9 |
| Sc | 21 | 27 | 67 | 6 | | | | 1.8 |
| Ti | 22 | 11 | 75 | 14 | | | | 2.1 |
| V | 23 | 8 | 71 | 20 | 1 | | | 2.1 |
| Cr | 24 | 10 | 68 | 21 | 1 | | | 2.1 |
| Mn | 25 | 49 | 50 | 1 | | | | 1.5 |
| Fe | 26 | 25 | 68 | 7 | | | | 1.8 |
| Co | 27 | 34 | 59 | 7 | | | | 1.7 |
| Ni | 28 | 30 | 64 | 6 | | | | 1.8 |
| Cu | 29 | 16 | 63 | 20 | 1 | | | 2.0 |
| Zn | 30 | 80 | 20 | | | | | 1.2 |
| Ge | 32 | 60 | 40 | * | | | | 1.4 |
| Sr | 38 | 2 | 98 | | | | | 2.0 |
| Y | 39 | 5 | 62 | 33 | | | | 2.3 |
| Zr | 40 | 1 | 47 | 45 | 7 | | | 2.6 |
| Nb | 41 | 1 | 24 | 51 | 22 | 2 | | 3.0 |
| Mo | 42 | 2 | 21 | 49 | 25 | 3 | | 3.1 |
| Rh | 45 | 46 | 43 | 10 | 1 | | | 1.7 |
| Pd | 46 | 23 | 67 | 9 | 1 | | | 1.9 |
| Ag | 47 | 13 | 61 | 25 | 1 | | | 2.1 |
| Cd | 48 | 68 | 32 | | | | | 1.3 |
| In | 49 | 66 | 34 | * | | | | 1.4 |
| Sn | 50 | 47 | 53 | | | | | 1.5 |
| Sb | 51 | 1 | * | | | | | 1.0 |
| Ba | 56 | | 100 | | | | | 2.0 |
| La | 57 | 1 | 76 | 23 | | | | 2.2 |
| Ce | 58 | 3 | 83 | 14 | | | | 2.1 |
| Pr | 59 | 3 | 69 | 28 | | | | 2.2 |
| Nd | 60 | | 83 | 17 | | | | 2.2 |
| Sm | 62 | 2 | 83 | 15 | | | | 2.1 |
| Gd | 64 | 2 | 76 | 22 | | | | 2.2 |
| Dy | 66 | 2 | 66 | 32 | | | | 2.3 |
| Ho | 67 | 2 | 66 | 32 | * | | | 2.3 |
| Er | 68 | 1 | 63 | 35 | 1 | | | 2.4 |
| Tm | 69 | 13 | 78 | 9 | | | | 2.0 |
| Yb | 70 | 3 | 88 | 8 | | | | 2.1 |
| Hf | 72 | 3 | 24 | 51 | 21 | 1 | | 2.9 |
| Ta | 73 | 2 | 33 | 38 | 24 | 3 | | 2.9 |
| W | 74 | 2 | 23 | 43 | 26 | 5 | 1 | 3.1 |
| Ir | 77 | 5 | 37 | 46 | 11 | 1 | | 2.7 |
| Pt | 78 | 12 | 69 | 18 | 1 | | | 2.1 |
| Au | 79 | 14 | 75 | 11 | | | | 2.0 |
| Pb | 82 | 36 | 64 | | | | | 1.6 |
| Bi | 83 | 83 | 17 | | | | | 1.2 |
| Th | 90 | | 24 | 64 | 12 | | | 2.9 |
| U | 92 | 20 | 40 | 32 | 8 | | | 2.3 |

Minimum beam noise is found empirically [31, 32] to occur when the extraction voltage and plasma density are optimally matched – the perveance match condition for optimum beam formation.

(By beam noise we mean the fractional fluctuation level of ion beam current about the mean beam current – $\delta i_b/i_b$.) Then the r.m.s. noise is a minimum of about 7%. Increased gas pressure in the discharge gap and in the extraction region improve the beam quality, but this method shifts the CSD to lower charge states. There is also a significant influence of magnetic field on beam noise and pulse stability.

Humphries and co-workers developed a technique for reduction in ion beam noise level by the use of wire meshes to reflect plasma electrons and to limit ion flow by its own space charge [33]. In the case of space-charge-limited flow, the extracted ion beam current is independent of the plasma fluctuation level and random plasma generation at the cathode spots. The method has been further developed specifically for accelerator injection application [34, 35].

Work carried out at GSI Darmstadt, Germany, over a period of several years has resulted in a vast improvement in both the beam noise and the beam pulse shape reproducibility [36]. By combining a number of techniques, including the use of meshes, addition of an optimized magnetic field, and geometric optimization, the performance of the GSI vacuum arc ion sources has been brought to the point where they are fully acceptable for routine accelerator injection application; the source is described below in section 5.2. The good performance of this source with respect to beam noise is evident from Fig. 6. Figure 6(a) shows the pulse shape of a uranium ion beam, pre-analysis, with all charge states, as measured by a Faraday cup 30 cm downstream: current is 156 mA and beam noise is ±4%. Figure 6(b) shows the post-analysis $U^{4+}$ beam as measured by a beam transformer at the entrance to the pre-linac radio-frequency quadrupole (RFQ) accelerator, a point about 12 m downstream from the ion source: $U^{4+}$ beam current is about 25 mA and beam noise is ±5%. The pulse shape, beam noise, and reproducibility are adequate for the source to be used routinely at GSI for high-current injection of metal ions, especially $U^{4+}$, into the UNILAC and SIS accelerators.

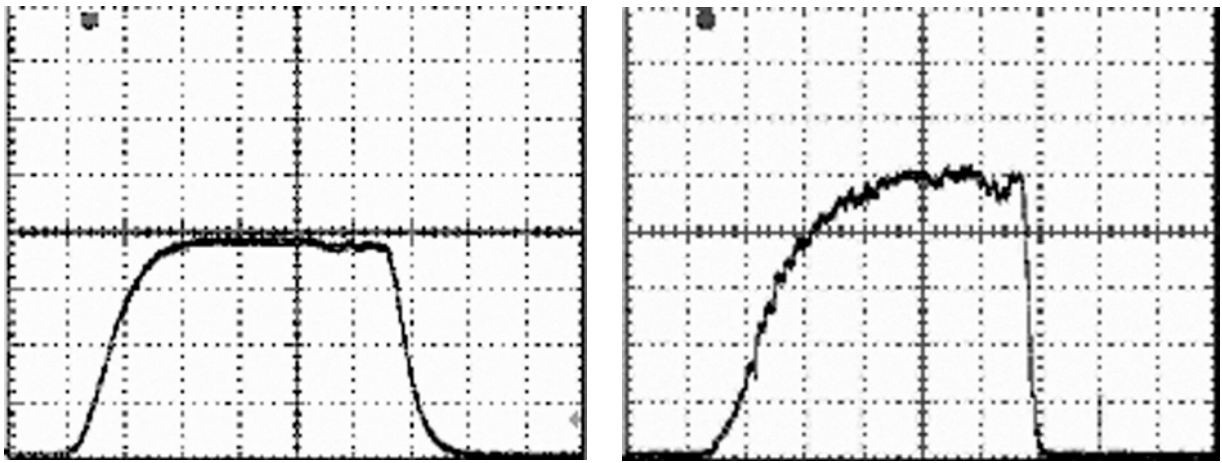

**Fig. 6**: (a) Uranium ion beam (all charge states) current pulse from the GSI vacuum arc source measured 30 cm from the source extractor; 40 mA/div. (b) Post-analysis $U^{4+}$ beam at the RFQ entrance, a distance of 12 m from the source; 5 mA/div. Sweep speed is 100 μs/div.

## 4.5 Gaseous operation

The vacuum arc ion source is above all a *metal* ion source. However, it can be operated so as to form beams that are controllable mixtures of gas and metal ions. Gas is fed into the source near the arc region where the plasma density is high, and a modest magnetic field ($B \sim 200$–400 G) is applied. The gas is ionized and gaseous ions are mixed with the metal plasma and extracted ion beam. The fraction of gaseous ions is regulated by the neutral gas feed, and the gas-to-metal ion ratio can be controlled from zero up to as much as about 99% gaseous. An example of this effect is shown in Fig. 7 for a titanium ion beam with added nitrogen. In this case a magnetic field was added to the arc region, and thus the low-pressure Ti charge states are somewhat elevated over the charge state distribution that is usual for the case of zero *B*. As the pressure is increased, there is a slow downshift of Ti states, and a simultaneous increase in the fraction of N ions in the beam. Two conclusions follow from these kinds of observations: (i) high fractions of highly charged ions call for the lowest possible background gas pressure; and (ii) ambient gas pressure provides a means for controlling the ion charge state distribution towards lower states. These effects can be used to advantage. For example, hybrid Ti–N ion beams have application for implantation of subsurface TiN layers, and similar buried ceramic layers can be formed from other species such as Al–O, Zr–O, etc. [37]. In work related to accelerator injection, a vacuum arc ion source was used to provide $Mg^+$ ions for injection into the GSI heavy-ion accelerator, having increased the ion source gas pressure so as to maximize the fraction of singly charged ions with respect to the doubly charged $Mg^{2+}$ fraction [38].

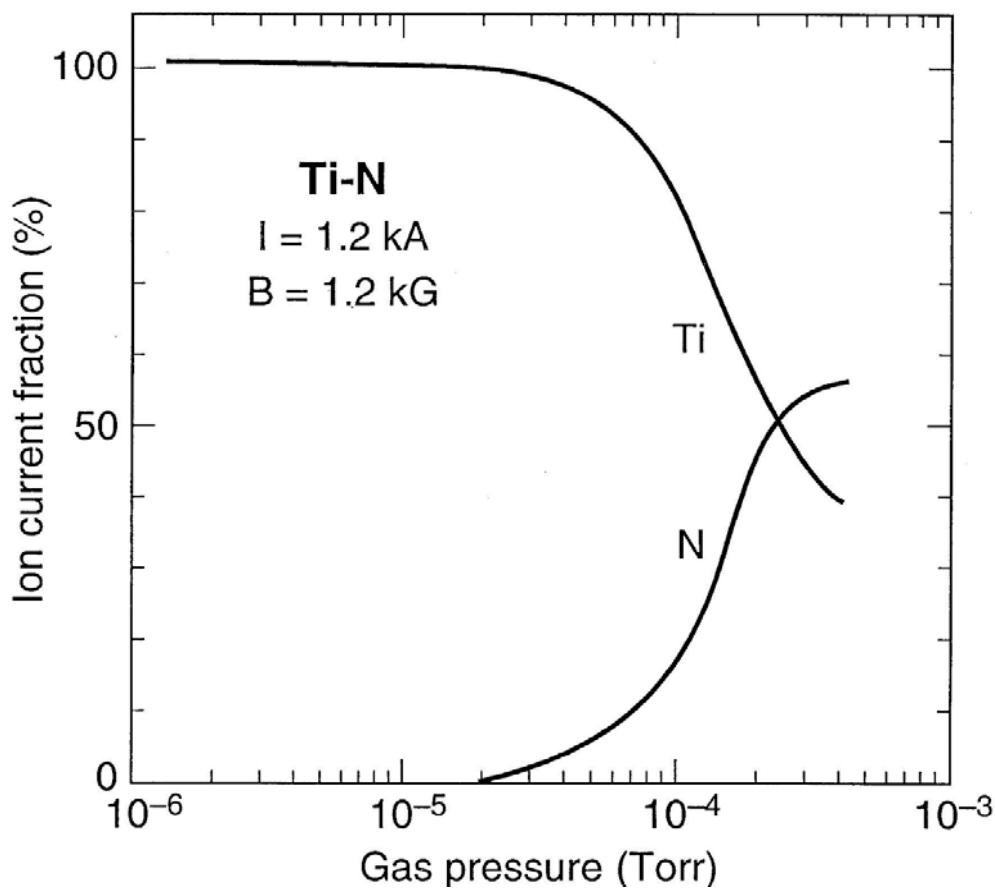

**Fig. 7**: Metal and gas ion beam composition fraction as a function of nitrogen gas pressure, for a hybrid Ti–N ion beam formed using the LBNL Mevva V ion source.

A way of operating a vacuum arc source in a 100% gaseous mode has been reported [39]. In this case, the vacuum arc feature is not used at all, but instead a hollow-cathode glow discharge is formed in the same ion source geometry.

There is a small gaseous contamination of the metal ion beam even at the lowest pressures at which the vacuum arc ion source is normally operated (~$10^{-6}$ Torr). The primary origin of this is condensation of gas (oxygen, nitrogen, water vapour) on the fresh front surface of the cathode from the residual ambient between arc pulses. This also affects, to some extent, the metal ion charge state distribution. The magnitude of the effect depends on the time between pulses, and can be minimized by operating at high repetition rate, greater than about 10 pps [40].

## 5 Source embodiments

We briefly describe here some of the vacuum arc ion sources that have been made and used at various laboratories. A comprehensive survey of all such sources is not attempted or implied. We distinguish the sources by the main application to which the work was directed.

### 5.1 Ion implantation application

The vacuum arc ion source has been adopted widely for metal ion implantation application, since the source can readily provide the high-current beams necessary for high-dose implantation for material surface modification. Vacuum arc ion implantation facilities have been set up at LBNL, Berkeley, USA [9, 10, 22, 23, 41–43], HCEI, Tomsk, Russia [11, 12, 44], NPI, Tomsk, Russia [14, 45, 46], Izmir, Turkey [47], the Australian Nuclear Science and Technology Organization (ANSTO), Sydney, Australia [48], Nagasaki, Japan [49], and most impressively at Beijing Normal University, China [50–52], and elsewhere.

The LBNL Mevva V ion source has operated steadily for many years with infrequent need for maintenance or repair. It remains a reliable workhorse and has served as a basic model for vacuum arc ion sources elsewhere. A schematic of the source is shown in Fig. 8 and a photograph in Fig. 9. It uses a 10 cm diameter, multi-aperture, accel–decel extractor configuration, incorporates a rotatable 18-cathode assembly, and can operate at up to about 65 kV extraction voltage or more (record high was 110 kV).

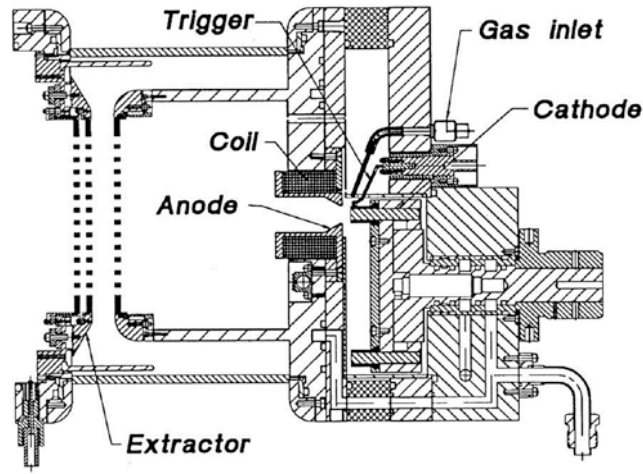

**Fig. 8:** LBNL Mevva V source, fitted with pulsed solenoid surrounding the arc region and with gas inlet feed

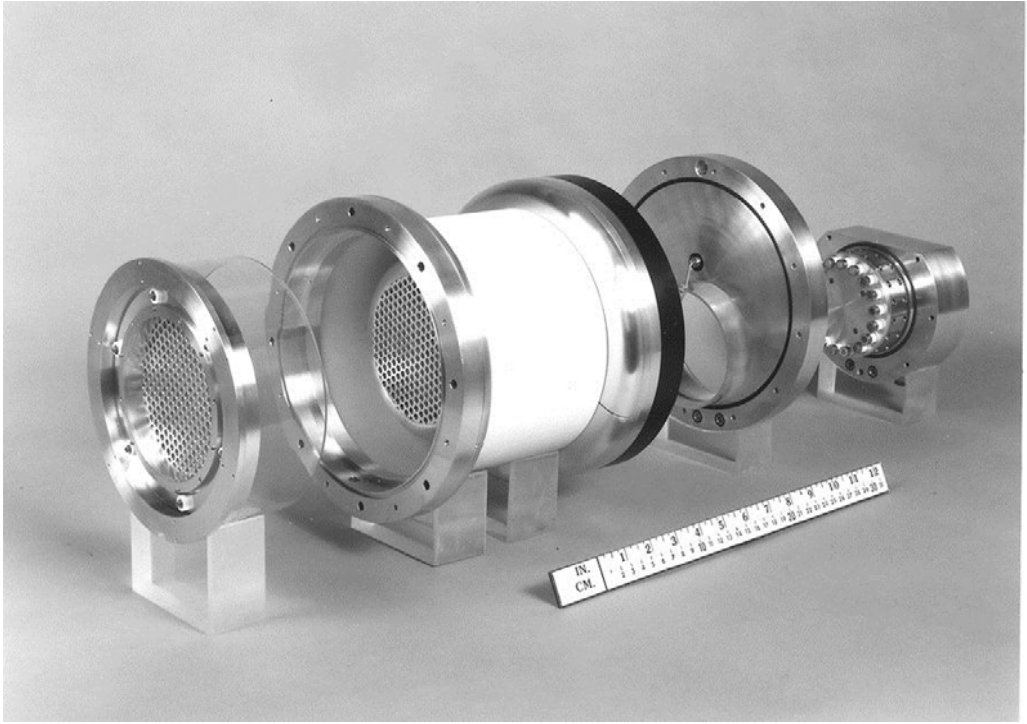

**Fig. 9**: LBNL Mevva V source, partly disassembled showing the 10 cm extractor grids on the left and the multiple cathode assembly on the right.

Beijing Normal University has a substantial vacuum arc ion source programme, initiated in 1987 with support from the National High Technology Research and Development Programme of China. Since then, more than 10 different embodiments of vacuum arc ion source-based implanters have been developed and deployed in factories and companies in China, Hong Kong, and Taiwan, with increasingly impressive ion source and implantation parameters. Among these, the largest presently in use is the Mevva 50 implanter, with a time-averaged ion beam current of 50 mA, installed at the Advanced Materials Center in the southern China city of Shenzhen. An even larger implanter has been made; this is the Mevva 100 implanter, with time-averaged ion beam current of 100 mA. This metal ion implantation facility is the largest vacuum arc metal ion implanter for industrial surface processing in the world.

## 5.2   Accelerator injection application

Vacuum arc ion sources have been used successfully at the Institute for Theoretical and Experimental Physics (ITEP), Moscow [53, 54] and at the GSI heavy-ion accelerator research centre at Darmstadt, Germany, for injecting metal ions into a linear accelerator and synchrotron, including uranium. At ITEP, beams of a range of metals were formed, including Be, C, Al, Fe, Cu, Zr, and W at beam current of up to 400 mA. For a pulse length of 100 μs and pulse repetition frequency 0.3 pps, operation times of up to 300–400 hours were obtained between necessary down-times for source maintenance (cathode replacement and other).

Much of the GSI work was performed collaboratively with the group at HCEI, Tomsk. For accelerator application, the quality of each individual ion beam pulse is important, as opposed to the case for ion implantation application. Beam noise and/or pulse shape variation from pulse to pulse can lead to poor accelerator performance. Thus the GSI source has incorporated a number of developments to enhance the high ion charge state fraction, to reduce the beam noise, and to improve the pulse-to-pulse beam shape reproducibility [34–36].

A schematic of the GSI source is shown in Fig. 10 and a photograph in Fig. 11. The cathode assembly holds 17 individual cathodes, each 17 mm long and 5.7 mm in diameter. The stainless-steel anode is located a distance of 15 mm from the cathode and has a central aperture 15 mm in diameter. Magnetic field coils are located external to the vacuum chamber. Two stainless-steel grids are installed to reduce the beam fluctuation level and to reduce plasma density. A multi-aperture accel–decel extraction system (13 holes, each 3 mm in diameter, aspect ratio 0.5) is used to form the ion beam. As an example of source performance, 25 mA of $U^{4+}$ ion current with an energy of 2.2 keV/u (156 mA full beam, or 170 mA cm$^{-2}$, electrical current) was measured at the entrance of the RFQ, for a typical extraction voltage of 32 kV, with a high fraction (67% electrical current fraction) of $U^{4+}$ ions, a post-analysis $U^{4+}$ beam noise level of about ±5%, and good beam pulse shape reproducibility. Examples of the total extracted uranium beam and the separated $U^{4+}$ beam pulse shapes are shown above in Fig. 6, and the main source and beam parameters are summarized in Table 2.

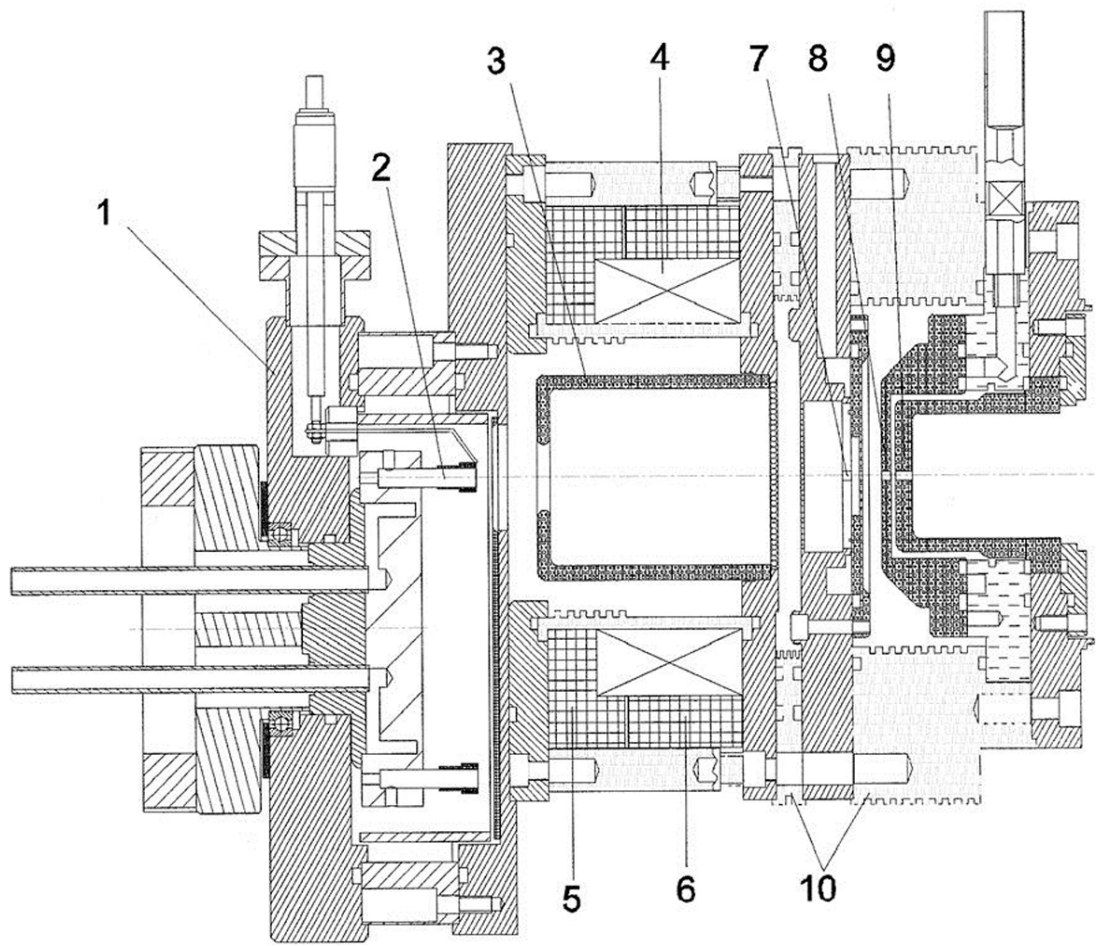

**Fig. 10**: GSI vacuum arc ion source: 1, cathode flange; 2, cathode; 3, anode (stainless steel); 4, SmCo cusp magnets (10 in total); 5, coil I; 6, coil II; 7, plasma electrode and grid; 8, screening electrode; 9, ground electrode; 10, insulators.

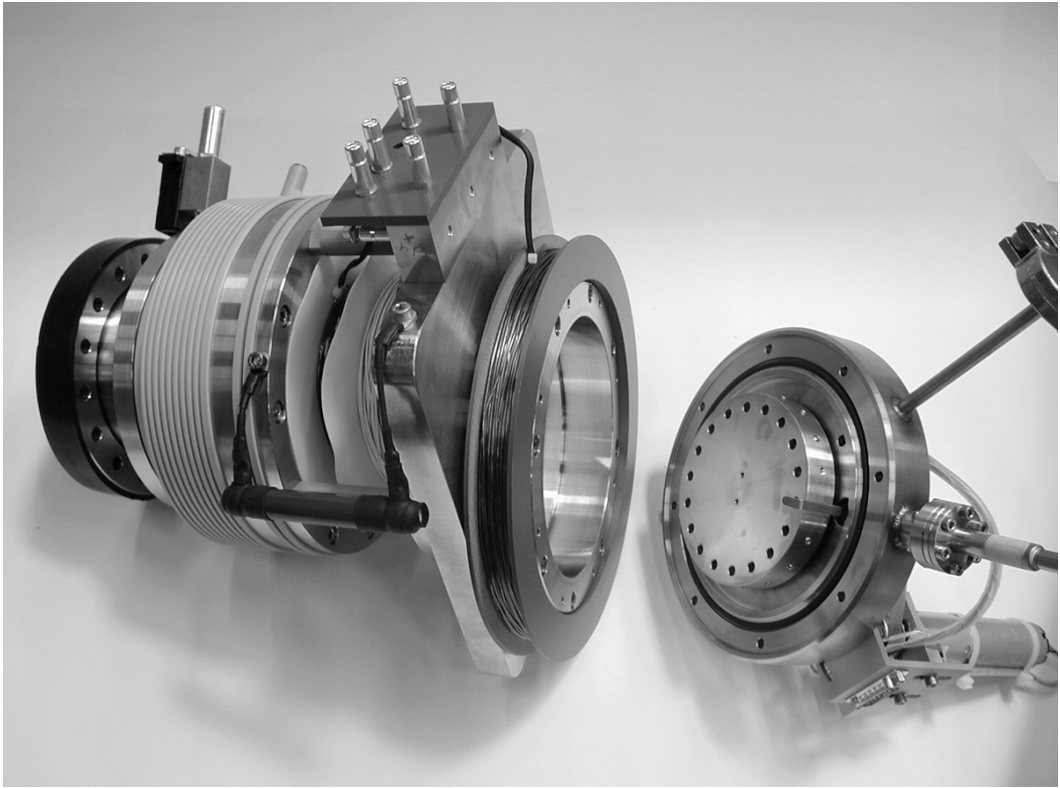

**Fig. 11:** Photograph of the partially disassembled GSI vacuum arc ion source

**Table 2:** GSI vacuum arc ion source and ion beam parameters.

| | |
|---|---|
| Ion species | Uranium |
| Ion charge state distribution | $U^{3+}$ = 16%, $U^{4+}$ = 67%, $U^{5+}$ = 14%, $U^{6+}$ = 3% |
| Extracted ion beam current | 156 mA at 35 kV |
| Accelerated ion beam current | 55 mA at 131 kV |
| Analysed $U^{4+}$ ion beam current | 25 mA |
| Pulse length / repetition rate | 0.6 ms / 1 pps |
| Extraction system | 13 × 3 mm, multi-aperture |
| $\varepsilon_{x,y}$ (156 mA at 35 kV) | 200 π mm rad |
| $\varepsilon_{x,y}$ (55 mA at 131 kV) | 350 π mm rad |
| $\varepsilon_{x,y}$ (after separation, 15 mA) | 100 π mm rad |
| Noise, full beam / analysed (r.m.s.) | < ±4% / ±5% |
| Pulse-to-pulse stability | Better than 80% |
| Voltage breakdowns (extractor grids) | two per day |
| Cathode lifetime | 12 hours at 0.6‰ duty |
| Lifetime of the ion source (between maintenance) | 7 days for SIS injection (0.2‰ duty) |

This ion source has proven its capability in extended tests at the GSI high-current injector and has been put into regular operation for the GSI accelerator facility, providing uranium ion beams with current about an order of magnitude greater than previously possible. We note that, because of the fixed input velocity required to match into the RFQ (2.2 keV/u), the ion source extraction voltage and the post-acceleration must be changed for the various different ion species used; thus each new species needs to be independently optimized. The GSI sources have been described in detail elsewhere [34−36, 55−57].

### 5.3 Heavy-ion fusion application

Heavy-ion fusion (HIF) is one approach to the problem of controlled thermonuclear power production, in which a small DT (deuterium-tritium) target is bombarded by an intense flux of heavy ions and compressed to fusion temperatures. There is a need in present HIF research and development for a reliable ion source for the production of heavy-ion beams with low emittance, low beam noise, ion charge states $Q$ = 1+ to 3+, beam current 0.5 A, pulse width 5–20 μs, and repetition rate 10 pulses per second. The suitability of a vacuum arc ion source for this application has been explored [58−61]. Energetic, high-current, gadolinium ion beams were produced, using the LBNL Mevva V source, with parameters as required or close to those required, and it is probable that the performance parameters can all be improved yet further in an optimized source design. An example of the beam pulse obtained is shown in Fig. 12, which shows an oscillogram of a Gd beam (85% in the $Gd^{2+}$ charge state) formed at an ion energy of 120 keV (60 kV extraction voltage), a collected beam current of 120 emA (electrical milliamperes), for a 20 μs pulse width and a pulse rise time of less than 1 μs [59]. The beam emittance was about 0.3 π mm mrad (normalized). These preliminary results indicate that a vacuum arc-based metal ion source of this kind is a good potential candidate for an HIF ion source.

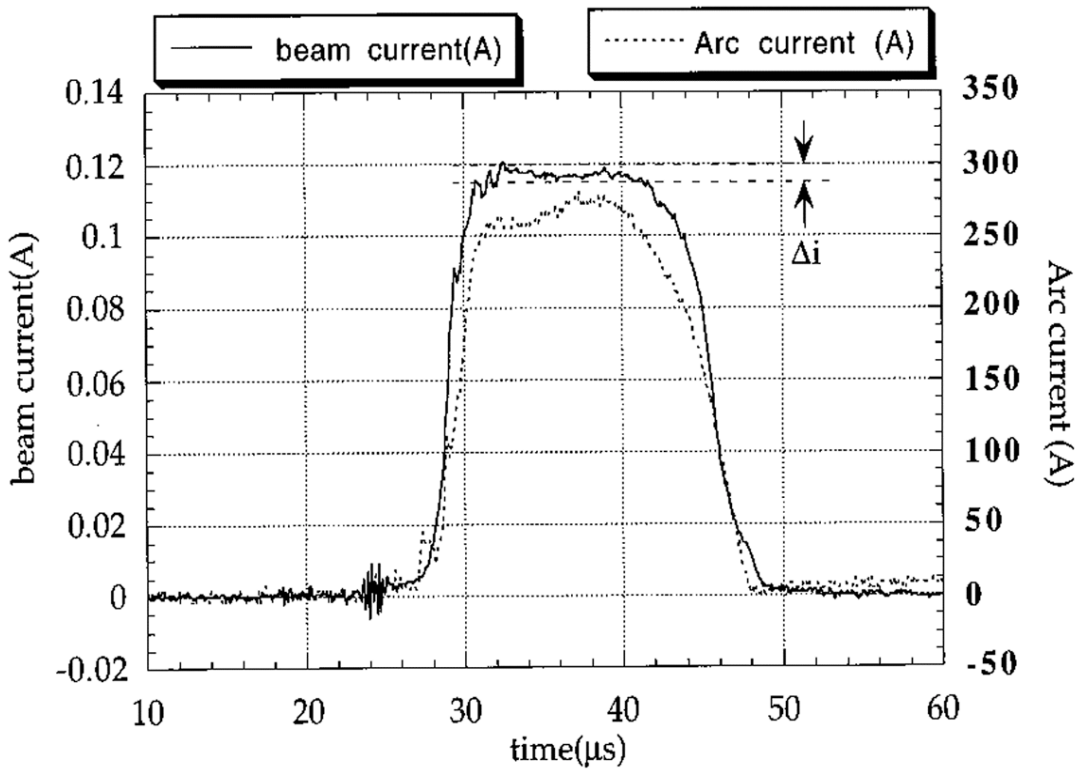

**Fig. 12**: Gd ion beam current (solid line), and arc current (dotted line), when the extraction is optimized (beam-formation perveance match condition).

### 6   Conclusion

High-current beams of metal ions can be provided by the vacuum arc ion source. Virtually all the metals of the Periodic Table, and carbon, have been used, and beams containing mixtures of metal ions and non-metallic species can also be made. The ions are in general multiply ionized but of low charge state, the beam typically containing a distribution of charge states spanning the rough range from 1+ to 5+, and techniques for increasing the charge states by as much as a factor of 2 have been demonstrated. The mean ion energy of the extracted beam is typically around 50–100 keV, with the charge state components of energy up to as much as several hundred kiloelectronvolts. The ion current that can be extracted is large, and the record high demonstrated, albeit with a 50 cm diameter extractor, is about 20 A; from smaller diameter extractors and thus for beams with a tighter emittance, beam current of 100–200 emA (all charge states) is typical. Although beam noise and shot-to-shot irreproducibility were initially a daunting concern, these features have been vastly improved, in particular by the GSI/HCEI collaborative work and as manifested by the highly successful use of the GSI vacuum arc ion source for routine production and accelerator injection, particularly of high-current $U^{4+}$ beams.